\documentclass{PoS}

\newcommand{\ApJ}{ApJ}

\newcommand{\etal}{et alii}

\newcommand{\AMS}{\textsf{AMS}}
\newcommand{\eg}{\textit{e.g.}} 
\newcommand{\ie}{\textit{i.e.}}

\newcommand{\citep}{\cite}
\newcommand{\citet}{\cite}

\newcommand{\Hyd}{\textrm{H}}
\newcommand{\He}{\textrm{He}}
\newcommand{\Li}{\textrm{Li}}
\newcommand{\Be}{\textrm{Be}}
\newcommand{\B}{\textrm{B}}
\newcommand{\C}{\textrm{C}}
\newcommand{\N}{\textrm{N}}
\newcommand{\Oxy}{\textrm{O}}
\newcommand{\Al}{\textrm{Al}}

\newcommand{\BC}{\textrm{B}/\textrm{C}}

\newcommand{\pbarp}{\textrm{\ensuremath{\bar{p}/p}}}
\newcommand{\pbar}{\textrm{\ensuremath{\bar{p}}}}
\newcommand{\nbar}{\textrm{\ensuremath{\bar{n}}}}
\newcommand{\dbar}{\ensuremath{\rm \overline{d}}}
\newcommand{\tbar}{\ensuremath{\rm \overline{t}}}
\newcommand{\hebar}{\ensuremath{\rm \overline{He}}}
\newcommand{\htwobar}{\ensuremath{\rm \overline{^{2}H}}} %2H
\newcommand{\hetbar}{\ensuremath{\rm \overline{^{3}He}}} %2H
\newcommand{\htbar}{\ensuremath{\rm \overline{^{3}H}}}
\newcommand{\abar}{\ensuremath{\rm \overline{A}}}

\newcommand{\Q}{\ensuremath{Q}}

\newcommand{\p}{\ensuremath{p}}
\newcommand{\n}{\ensuremath{n}}
\newcommand{\pp}{\ensuremath{p}-\ensuremath{p}}
\newcommand{\pC}{\ensuremath{p}-\ensuremath{C}}
\newcommand{\captionsize}{\footnotesize}
\usepackage{xspace}

\title{Production of antimatter nuclei\\ in Galactic cosmic rays}
\ShortTitle{Production of antimatter nuclei in Galactic cosmic rays}

\author{\speaker{Alberto Oliva}\thanks{E-mail: {alberto.oliva@cern.ch}}\\
Centro de Investigaciones Energ{\'e}ticas, Medioambientales y Tecnol{\'o}gicas -- CIEMAT, E-28040 Madrid, Spain\\
}

\author{{Nicola Tomassetti}\thanks{E-mail: {nicola.tomassetti@cern.ch}}\\
Department of Physics and Earth's Science, Universit{\`a} di Perugia, and INFN-Perugia, I-06100 Perugia, Italy\\
}

\author{{Jie Feng}\thanks{E-mail: {jie.feng@cern.ch}}\\
  Massachusetts Institute of Technology -- MIT 02139, Cambridge, Massachusetts, USA
}

\abstract{
Antimatter nuclei in cosmic rays (CRs) are a promising tool for the indirect detection of dark-matter annihilation signatures.
However, the search of new-physics signals in CRs relies on our knowledge of the astrophysical 
antimatter background which, 
in turns, depends critically on the several fragmentation cross-sections that regulate production and destruction 
of antiparticles in the interstellar medium.
In this work, we have re-evaluated the astrophysical background of CR antiproton, antineutron, 
and antihelium nuclei in Galactic CRs using improved calculations. 
The production cross-sections of individual antinucleons are constrained using updated
calculations that make use of recent accelerator data. 
The production of antideuteron and antihelium nuclei is calculated using an improved model of 
nuclear coalescence that accounts for the asymmetry in antineutron and antiproton production.
We discuss the cross-section induced uncertainties and show that they are dominating
in comparison with other uncertainties of astrophysical origin.
}

\FullConference{35th International Cosmic Ray Conference -- ICRC2017 --\\
		10-20 July, 2017\\
		Bexco, Busan, Korea}

\begin{document}

%%%%%%%%%%%%%%%%%%%%%%%%%%%%%%%%%
\section{Introduction}
\label{Sec::Introduction}
%%%%%%%%%%%%%%%%%%%%%%%%%%%%%%%%%

Antimatter nuclei in cosmic rays (CR) represent a promising discovery channel for the indirect search of dark matter.
Annihilation of decay products of dark matter particles in the Galaxy may generate antiprotons (\pbar) and antineutrons (\nbar) which,
in turn, can merge into light antinuclei such as antideuteron \htwobar{} (or \dbar),  antihelium \hetbar{} (\emph{tout court} \hebar), or 
antitritium \htbar{} (or \tbar), where the latter decays rapidly into \hebar{} \citep{Aramaki2016,Cirelli2014}
Antiprotons have been detected since long time in CRs, and it
is generally agreed that the vast majority of these particles originates from secondary production mechanisms, \ie, 
from collisions of high-energy protons and nuclei with the gas nuclei of the interstellar medium (ISM).
The antiproton/proton (\pbarp) ratio CRs has been recently measured with high precision 
by the Alpha Magnetic Spectrometer (\AMS) from 0.5 to 450\,GeV of kinetic energy \citep{Aguilar2016PbarP}. 
This ratio is found to be unexpectedly constant at $E\gtrsim$60\,GeV, in contrast with
standard model predictions and with the trend suggested by the \BC{} ratio \citep{Aguilar2016BC}.
On the other hand, the physics of CR propagation is poorly understood and new physics
scenarios have been proposed \citep{TomassettiDonato2015,PbarSnr}.
CR propagation models also suffer from large uncertainties in the predition of the \pbarp{} ratio \citep{Giesen2015,Feng2016}. % ZZZZZ \citep{}.
In contrast to antiprotons, heavier antinuclei have never been observed in CRs. Their detection, however,
have chances to be achieved by the next generation of antimatter experiments in space.
For these nuclei, the astrophysical background is expetted to be kinematically suppressed at sub-GeV/n energies,
while DM signals peak in this energy window according to several models of DM annihilation.
In this paper, we report calculations of secondary antinuclei fluxes and their uncertainties.
In particular we focus on nuclear-physics calculations, and associated uncertainties,
for the production antinucleons and antinuclei arising from collisions of high-energy CR nuclei with the ISM.

%%%%%%%%%%%%%%%%%%%%%%%%%%%%%%%%%%%%%%%%%%%%%%%%%%%%%%%
\section{Calculations}
\label{Sec::Calculations}
%%%%%%%%%%%%%%%%%%%%%%%%%%%%%%%%%%%%%%%%%%%%%%%%%%%%%%%

Calculations of CR fluxes account for the acceleration and diffusive propagation of CRs in the turbulent magnetic fields, 
during which secondary particles and antiparticles are generated. 
\textit{Primary} CRs such as \p, \He, or \C-\N-\Oxy{} nuclei are believed to be accelerated inside supernova remnants (SNRs)
to power-law energy spectra of the type $\Q^{\rm pri}\sim\,E^{-\nu}$, with $\nu\approx$\,2.1-2.4. 
\textit{Secondary} particles such as \Li-\Be-\B{} nuclei or \pbar-\dbar-\hebar{} antinuclei
are mostly generated by collisions of primary nuclei with the gas of the ISM. 
\textit{Secondary-to-primary} ratios of stable nuclei, and in particular the \BC{} ratio, are used to constrain the
CR propagation parameters and to eventually predict the near-Earth fluxes of secondary antiparticles \citep{Grenier2015}.
In propagation calculations, the generic source-term for a $j$-type CR particles is represented by a ${\Q^{j}}$-function
which is a combination of the type ${\Q}^{j}={\Q}^{j}_{\rm pri} + {\Q}^{j}_{\rm sec}$, \ie, including 
for \emph{primary} CRs accelerated in SNRs and \emph{secondary} fragmentation/decay-induced terms from $k$-type species.
The primary (SNR) term is generally modeled by spatial and energy dependent parametric source functions, 
${\Q}_{\rm pri}^{j}=2h\delta(z)f(r)q^{j}_{0}s^{j}(E)$.
For  $k\rightarrow j$ decay of radioactive nuclei with corresponding lifetime $\tau_{0}^{k\rightarrow j}$, 
the source term is of the type as ${\Q}^{j}=\sum_{k} \N^{k}({\bf r},E)/(\gamma\tau_{0}^{k\rightarrow j})$.
For $k\rightarrow j$ nuclear fragmentation processes, the general form is:
\begin{equation}\label{Eq::GeneralSourceTerm}
{\Q}^{j}_{\rm sec}(E) = \sum_{k>j}  \int_{0}^{\infty} dE^{k} \N^{k}(E^{k})  \sum_{i} \Gamma^{k\rightarrow j}_{i}(E,E^{k}) 
\end{equation}
where $\Gamma^{k\rightarrow j}_{i}$ is the differential rate of $j$ particle production 
with kinetic energy $E$ from collisions of $k$-type CR particles, with density $\N^{k}$, with $i$-type targets of the ISM component (with number density $n_{i}$).
These interactions couple the equations of every $j$-type nucleus to those of all heaveier $k$-nuclei.
The system is resolved by starting from the heavier nucleus which is assumed to be purely primary. 
In practice, the bulk of antimatter in the Galaxy produced by the dominant CR components, \ie, protons and alpha CR particles,
colliding with hydrogen and helium atoms of the ISM. Thus the dominant CR-ISM collisions processes are \p-\Hyd, \p-\He, \He-\Hyd, and \He-\He.

%%%%%%%%%%%%%%%%%%%%%%%%%%%%%%%%%%%%%%%%%%%%%%%%%%%%%%%
\subsection{Antiproton production}
\label{Sec::Pbar}
%%%%%%%%%%%%%%%%%%%%%%%%%%%%%%%%%%%%%%%%%%%%%%%%%%%%%%%
%
For proton-proton collisions, the typical antiproton production process is $p+p \rightarrow \bar{p} + X$.
The corresponding differential cross-section can be described in terms of the Lorentz-invariant distribution function:
\begin{equation}
f(k+i \rightarrow \bar{p} + X) = E_{c}\frac{d^{3}\sigma}{dp^{3}_{\bar{p}}} 
\end{equation}
The total differential cross section is given by the integral over the angle between the incoming proton and the direction of the ejected antiproton.
In the past 20 years, antiproton production cross-sections parameterised by \citet{TanNg:1983} have been widely used in CR propagation calculations. 
Due to the lack of high-energy measurements, this semi-empirical parameterisation has been tuned to the experimental data available at the epoch
and then extrapolated to the relevant energies. Recently, high-energy collision experiments have triggered some 
efforts in updating the antiproton cross-section models \cite{diMauro:2014zea,Kachelriess:2015wpa,Kappl:2014}. 
%%%%%%%%%%%%%%%%%%%%%%%%%%%%%%%%%%%%%%%%%%%%%%%%%%%%%%%
\begin{figure}[!t]
\includegraphics[width=0.46\textwidth]{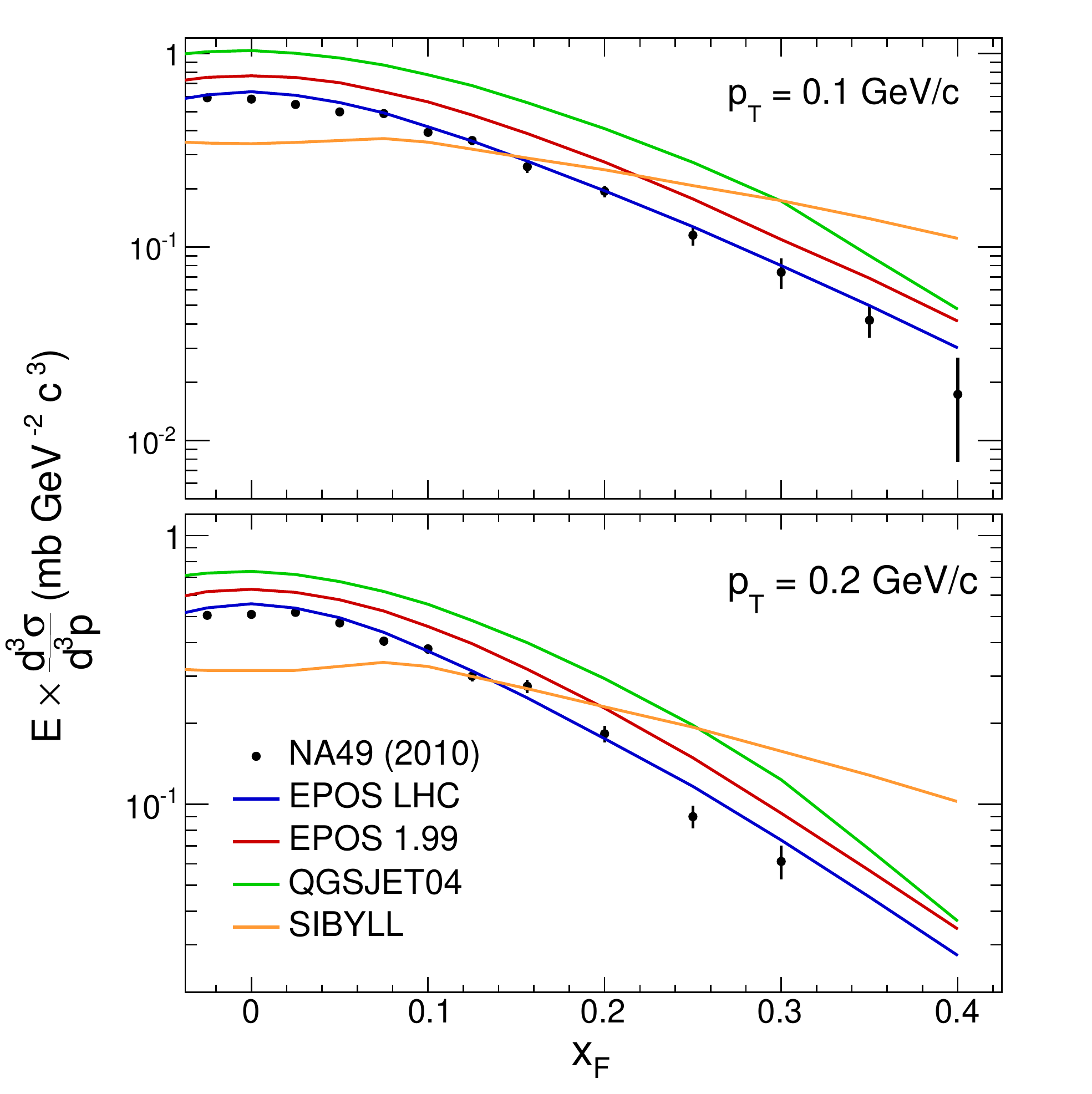}
\qquad
\includegraphics[width=0.46\textwidth]{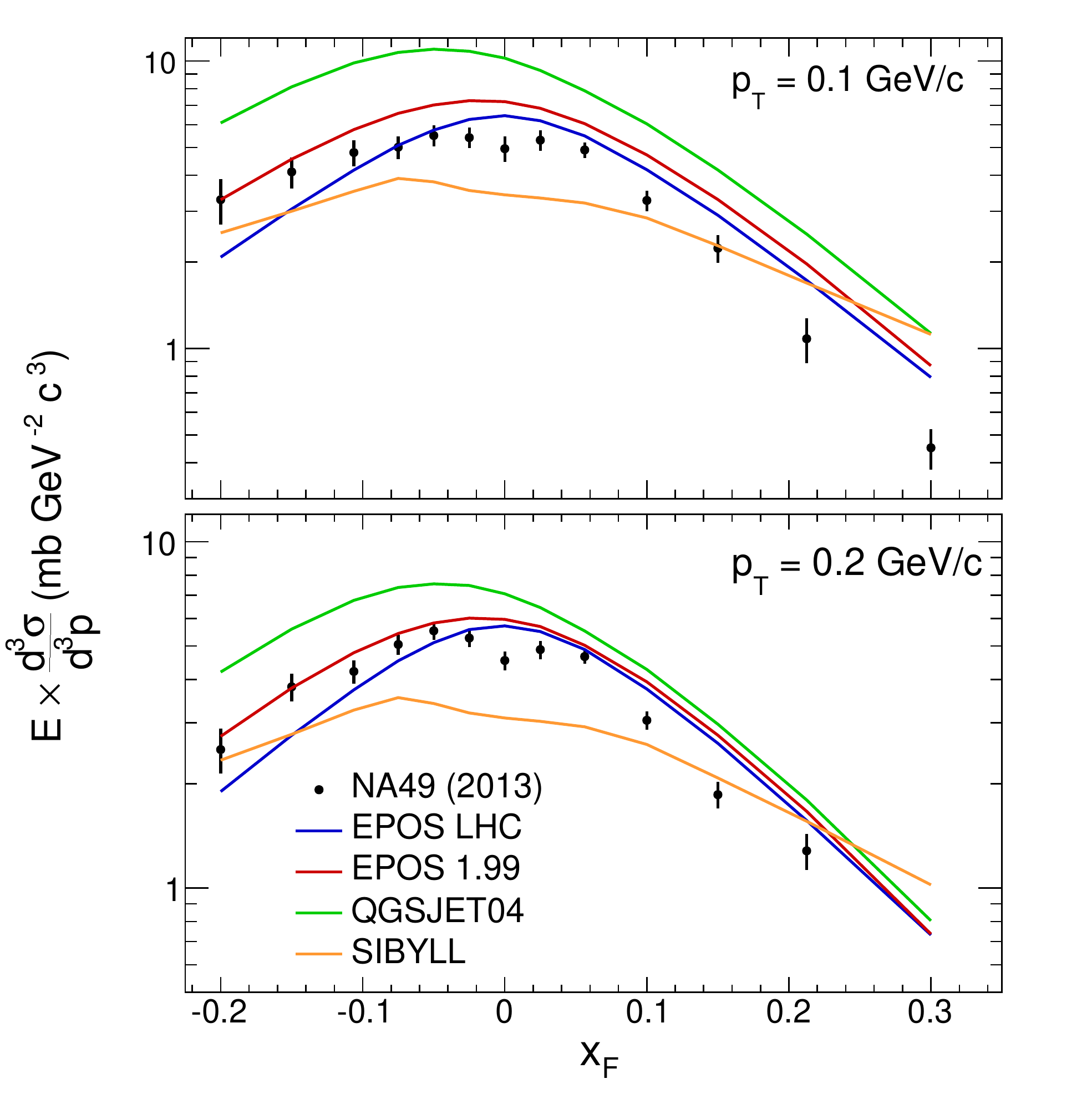}
\caption{
  Invariant cross-section for \pp{}$\rightarrow$\,\pbar{} (left) and \pC{}$\rightarrow$\,\pbar{} (right) production measured by the NA49 experiment~\cite{NA49} (dots) 
for two transverse momenta  $p_{T} = 0.1, 0.2$ GeV/c (top, bottom), as function of the Feynman-$x$ variable $x_{F}$. Corresponding calculation from MC generators are superimposed (lines).
}
\label{Fig::na49_pp}\label{Fig::na49_pC}
\end{figure}
%%%%%%%%%%%%%%%%%%%%%%%%%%%%%%%%%%%%%%%%%%%%%%%%%%%%%%%
We have made use of recent data on antiproton production from \pp{} and \pC{}  collisions reported by
NA49\,\cite{NA49}, BRAHMS\,\cite{Arsene:2007jd}, and ALICE\,\cite{Aamodt:2011zj}, to constrain the cross-section
calculations of MC generators such as
\texttt{EPOS LHC}, \texttt{EPOS 1.99}~\cite{Pierog:2013ria}, \texttt{SIBYLL}~\cite{Engel:1999db}, and \texttt{QGSJET-II-04}~\cite{Ostapchenko:2010vb}.
These MC generators are widely used in the simulation of extensive CR air showers and have
been recently tuned to reproduce minimum bias LHC Run-1 data.
%%%%%%%%%%%%%%%%%%%%%%%%%%%%%%%%%%%%%%%%%%%%%%%%%%%%%%%

Figure\,\ref{Fig::na49_pp} shows the \pbar{} production cross-sections measured by the NA49 experiment at CERN in \pp{} (left) and \pC{} (right) collisions.
In this experiment, antiprotons are generated by a 158 GeV/c momentum proton beam extracted at the Super Proton Synchrotron interacting on H or C steady targets.
In the figures, the $\bar{p}$ production invariant cross-section is presented for two different transverse momenta, $p_{T}=$\,0.1 and 0.2 GeV/c.
Corresponding calculation from MC generators are superimposed. From the comparisons with the data, it can be seen that \texttt{EPOS LHC} performs slightly better than other models. 
%%%%%%%%%%%%%%%%%%%%%%%%%%%%%%%%%%%%%%%%%%%%%%%%%%%%%%%
\begin{figure}[!t]
  \includegraphics[width=0.45\textwidth]{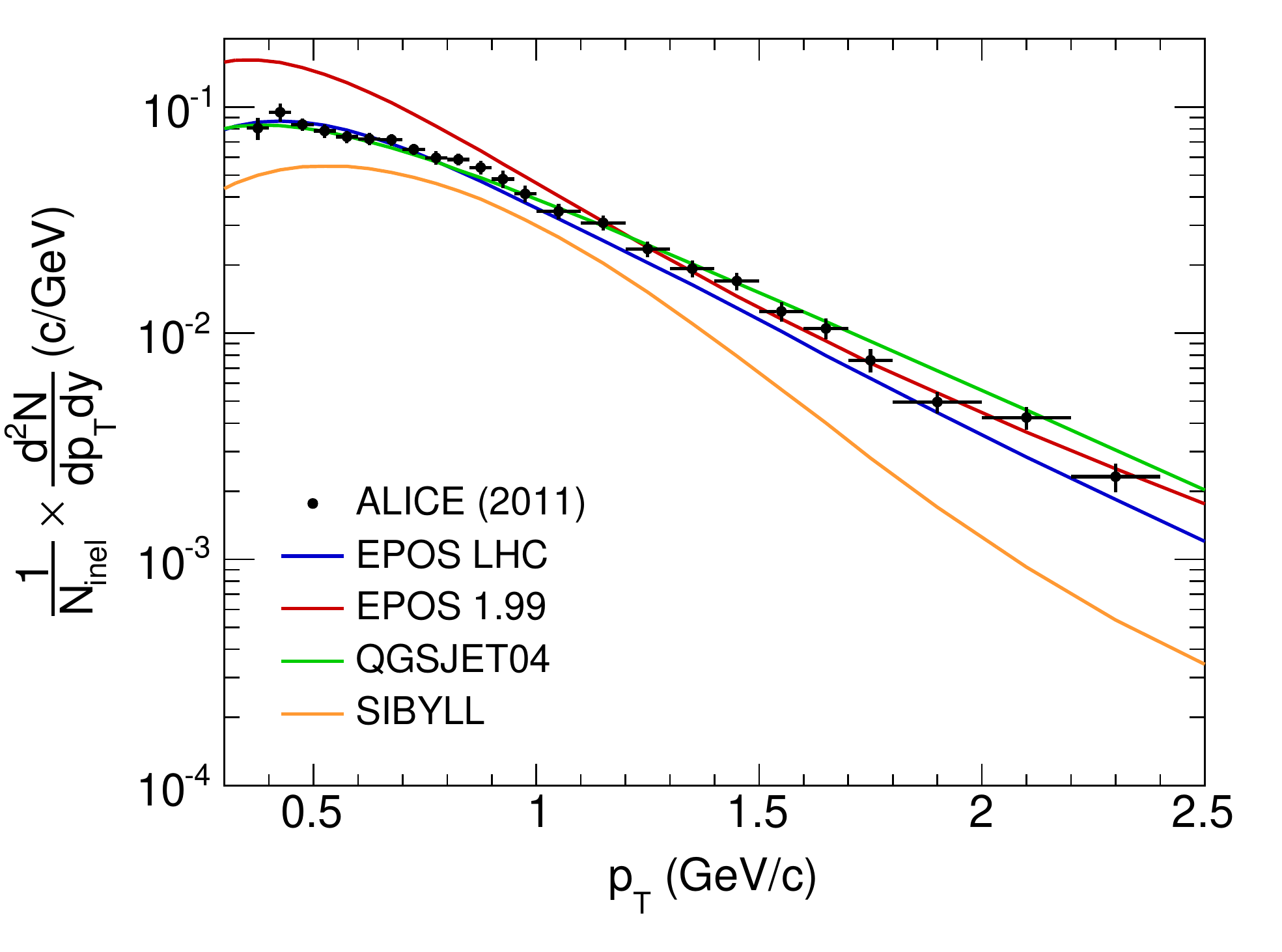}
\qquad
  \includegraphics[width=0.48\textwidth]{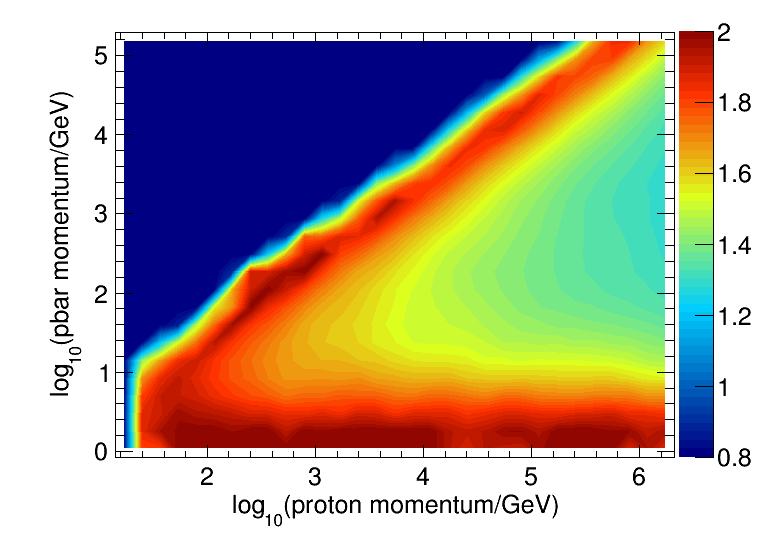} 
\caption{\captionsize%
  Left:
  the \pp{}$\rightarrow$\,\pbar{} production yields measured by ALICE\,\cite{Aamodt:2011zj} (dots) 
  for rapidity in the range $\left|y\right| < 0.5$ as function of $p_{T}$. 
  Calculation from MC generators are superimposed (lines). 
  Right:
  mean \nbar/\pbar{} ratio from \pp{} collisions simulated with \texttt{EPOS LHC} as function of the 
  primary and secondary proton momenta $p_{\rm pri}$ and $p_{\rm sec}$
}
\label{Fig::alice_pp}\label{Fig::ccNbarPbar} 
\end{figure}
%%%%%%%%%%%%%%%%%%%%%%%%%%%%%%%%%%%%%%%%%%%%%%%%%%%%%%%
Other comparisons have been done with \pp{} measurements reported by the BRAMHS collaborations
at the RHIC collider in Brookhaven \citep{Feng2016}.
In Fig.\,\ref{Fig::alice_pp} (left) it is shown the \pbar{} production yield measured by the ALICE experiment at CERN using \pp{} interactions at LHC with $\sqrt{s} = 900$\,GeV.
Data are shown for the rapidity range $\left|y\right|$ $<$ 0.5 as function of $p_{T}$. The comparison with MC generators shows a good agreement
with \texttt{EPOS LHC} and  \texttt{QGSJET-II-04} at all $p_{T}$-values, and agreement with \texttt{EPOS 1.99} for $p_{T} > 1.2$ GeV/c.

%%%%%%%%%%%%%%%%%%%%%%%%%%%%%%%%%%%%%%%%%%%%%%%%%%%%%%%
\subsection{Antineutron production}
\label{Sec::Nbar}
%%%%%%%%%%%%%%%%%%%%%%%%%%%%%%%%%%%%%%%%%%%%%%%%%%%%%%%

Another important physics input for the antiproton flux calculation is the production of antineutrons.
Antineutrons in fact decay rapidly into antiprotons, with a rest lifetime $\tau_{0}\approx$\,15\,min,
therefore contributing appreciably to the observed antiproton flux.
Traditionally it is assumed that $\sigma_{pp\rightarrow\bar{n}} \equiv \sigma_{pp\rightarrow \bar{p}}$,
which gives a \nbar-contribution to the flux identical to that of direct \pbar-contribution. 
However, preliminary measurements on deuteron-proton interactions have suggested a slight enhancement of
the antiproton yield from \n-\p{} collisions with respect to \pp{} \citep{Fisher2003},
which reveals a preferential production of \p-\nbar{} pairs compared to \n-\pbar{} pairs generated in \pp{} collisions.
Based on these data, like in other recent studies, we introduced a factorized scaling of the type
$\sigma_{pp\rightarrow\bar{n}} \equiv \kappa_{n} \times \sigma_{pp\rightarrow \bar{p}}$,
with $\kappa_{n}= 1.3\pm\,0.2$ \citep{diMauro:2014zea,Kappl:2014}.
Clearly, the use of a constant $\kappa_{n}$ over the whole phase space may be consider an over-simplification of the problem.
For instance, under \texttt{EPOS-LHC}, the multiplicity ratio \nbar/\pbar{} is found to vary from 1 to 1.9.
We also  account for uncertainties in \nbar-production, by assuming full correlation of \nbar-uncertainties with those from \pbar-production.
This contributes appreciably to the total errors on the CR antiproton source term.
Improving laboratory measurements on \nbar{} production is of paramount importance
for reducing the uncertainties in the prediction of the CR antiproton flux.

%%%%%%%%%%%%%%%%%%%%%%%%%%%%%%%%%%%%%%%%%%%%%%%%%%%%%%%
\subsection{Antideuteron and antihelium production} %%%
\label{Sec::Abar}                                   %%%
%%%%%%%%%%%%%%%%%%%%%%%%%%%%%%%%%%%%%%%%%%%%%%%%%%%%%%%

\begin{figure}[t]
\centering
\includegraphics[width=0.96\textwidth]{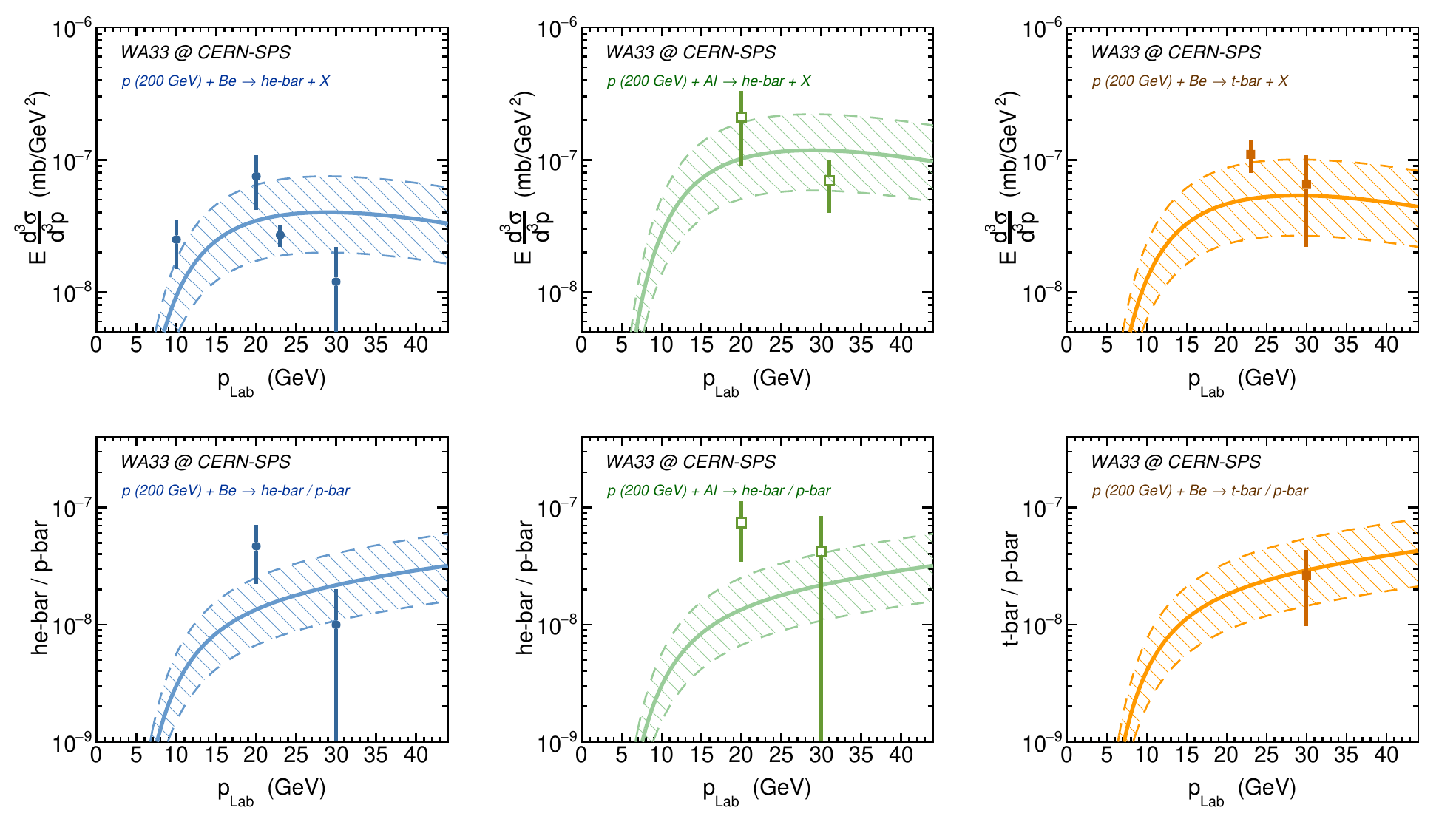} 
\caption{\captionsize% 
  Production cross sections for \hebar{} and \tbar{} from \p-\Be{} and \p-\Al{} collisions
  in comparison with calculations.
} 
\label{Fig::ccHebarXS} 
\end{figure}

Light antinuclei are formed by the fusion of antiprotons and antineutrons:
\dbar=\{\nbar+\pbar\}, \tbar=\{\nbar+\nbar+\pbar\}, and \hebar=\{\nbar+\pbar+\pbar\}.
Within the so-called coalescence model, calculations of antideuteron production were performed by several authors \citep{Donato2008}.
In the present work, we have devised an improved version of the standard coalescence model, which estimate the invariant differential cross 
section for the production of antimatter nuclei, 
by taking account for possible asymmetries in antiproton and antineutron production.
For the coalescence of $A$ antinucleons, and in particular  $Z$ antiprotons and $A-Z$ antineutrons, the generalized formula reads:
\begin{equation} \label{Eq::Coalescence}
 E_{\bar{A}} \frac{d^{3}N_{\bar{N}}}{dp^{3}_{\bar{N}}} = 
B_{\bar{A}} \times \left( E_{p} \frac{ d^{3}N_{p}}{dp^{3}_{p}} \right)^{Z}  \times \left( E_{n} \frac{ d^{3}N_{n}}{dp^{3}_{n}} \right)^{A-Z}\,,
\end{equation}
where $N_{\bar{A}}$ and $N_{p}$ and $N_{n}$ are the A-antinucleus, antiproton, and antinucleon production multiplicities, respectively.
These multiplicities are calculated following antiproton production parametrisation of \cite{diMauro:2014zea}.
The coefficient $B_{\bar{A}}$ is defined as $B_{\bar{A}}=\left(\frac{4\pi}{3}p_{0}^{3}\right)^{A-1}\frac{m_{\bar{A}}}{m^{A}_{p}}$. 
To include the production threshold effect we used the approach of Chardonnet et al. \citep{Donato2008}, Ansatz (1).
In comparison to previous works, however, our Eq.\,\ref{Eq::Coalescence} has been generalized for the case of asymmetric \nbar/\pbar{} production.
In light of the recent NA49 data \citep{Fisher2003}, as discussed in Sect.\,\ref{Sec::Nbar}, we have allowed for a $30\,\%$ asymmetry between \nbar{} and \pbar{} cross-sections.
This asymmetry has no impact in the coalescence calculations for \dbar{} particles, as the model is eventually constrained against the data.
For $A=3$ particles, however, we predict  a larger production of \htbar{} triplets (\pbar,\nbar,\nbar) with respect to \hetbar{} triplets (\pbar,\pbar,\nbar).
Since the former decay rapidly into \hebar, the total \hebar{} flux expected near-Earth collects both contributions, where the \htbar{} contribution is slighthy larger
than the ``direct'' \hebar{} production channel.
Finally, we have also accounted for destruction of secondary particles and for their subsequent production of tertiaries,
including non-annihilating reactions such as \abar+\p+\abar$^{\prime}$+X, where the
final state antiparticles have smaller energies \citep{Donato2008,InelasticXS,TanNg:1983}.
Further details on these calculations are left to a forthcoming work.
%%%%%%%%%%%%%%%%%%%%%%%%%%%%%%%%%%%%%%%%%%%%%%%%%%%%%%% 
\begin{figure}[t] 
\includegraphics[width=0.46\textwidth]{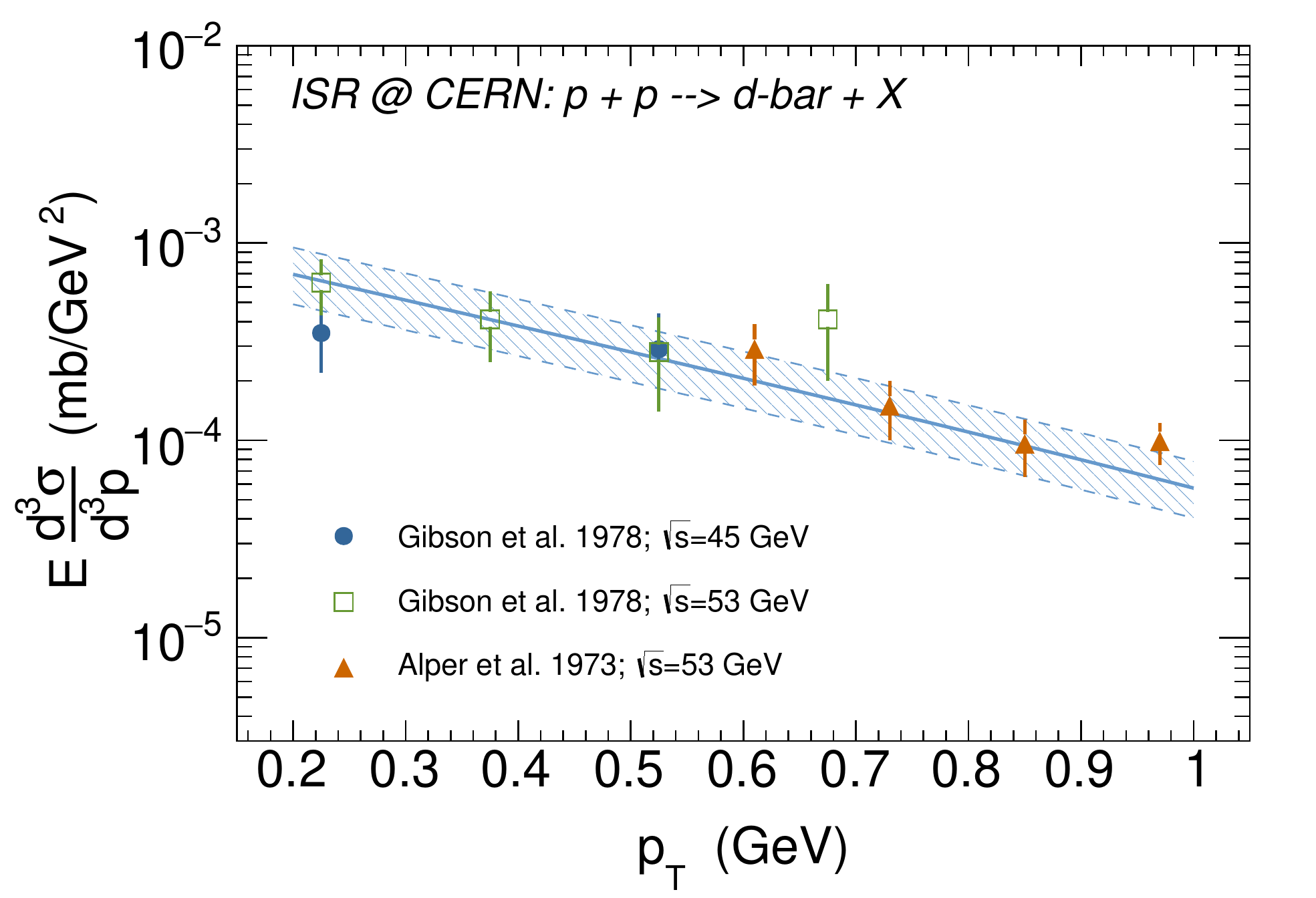} 
\qquad
\includegraphics[width=0.46\textwidth]{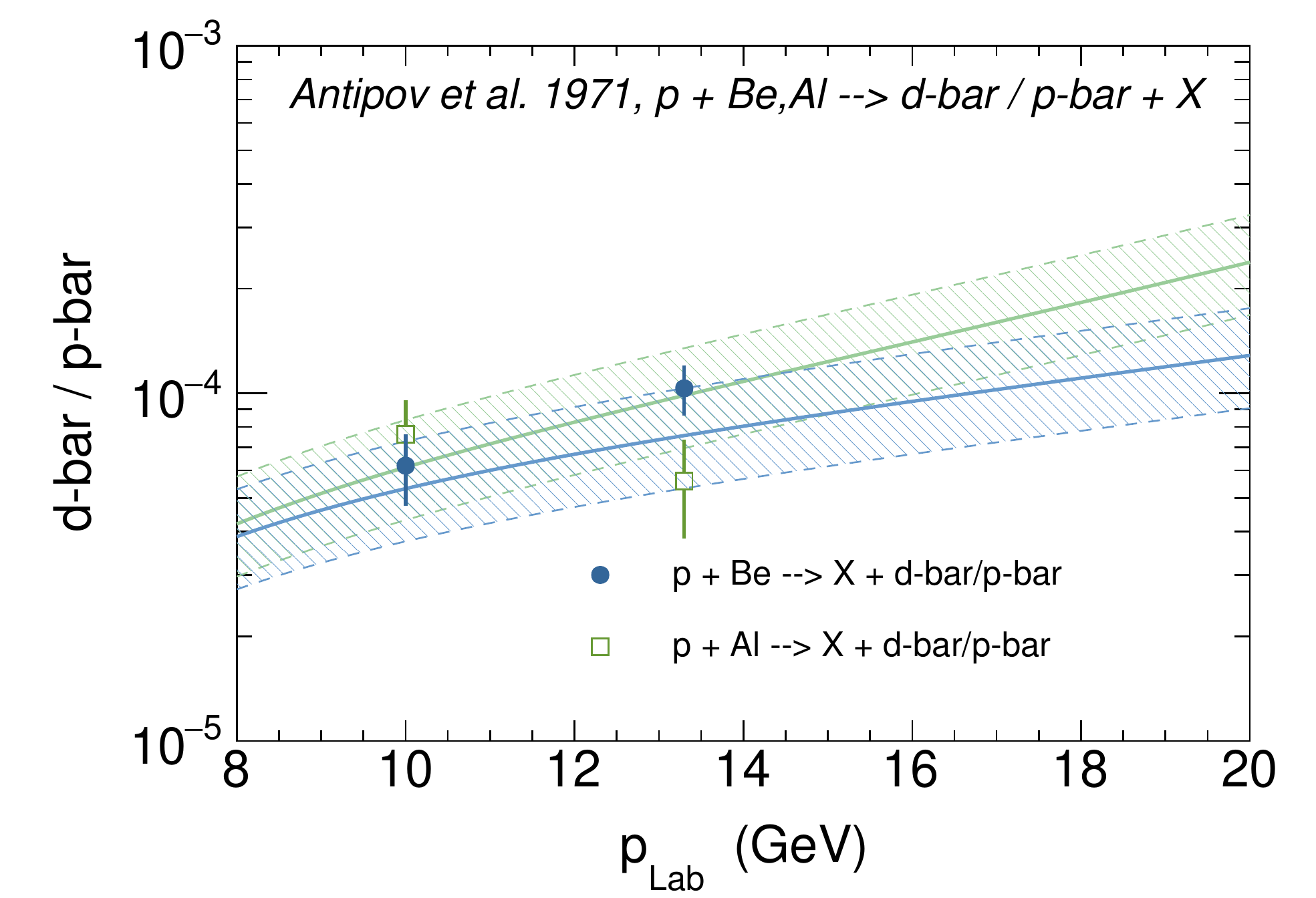} 
\caption{\captionsize% 
  Cross-section calculations for \dbar{} production (left) and \dbar/\pbar{} ratio (right)
  in comparison with accelerator data from ISR at CERN \citep{DbarXS}.
} 
\label{Fig::ccDbarXS} 
\end{figure} 
%%%%%%%%%%%%%%%%%%%%%%%%%%%%%%%%%%%%%%%%%%%%%%%%%%%%%%% 
We have made use of accelerator data to constrain the coalescence momentum parameter.
In Fig.\,\ref{Fig::ccDbarXS}, cross-section calculations for \dbar{} production are shown in comparison with accelerator data from ISR at CERN, Geneve  \citep{DbarXS}.
The data enabled us to infer the parameter $p_{0}$ and its uncertainty.  
In the right panel, data on \dbar/\pbar{} cross-section ratio are shown for \p+\Be{} and \p+\Al{} collisions as function of the proton momentum.
From the data, we set the coalescence momentum to $p_{0}\cong 90$\,MeV/c, with a 12\,\% uncertainty which is represented in the figures as shaded band.
From this setting, we compute the production cross-sections of \hebar{} and \tbar{} triplets, shown in Fig.\,\ref{Fig::ccHebarXS}
in comparison with the data. In the bottom panels of the figure, branching ratios \hebar/\pbar{} and \tbar/\pbar{} are shown.
Calculations are based on the same coalescence momentum used for describing \dbar-production.
Within the uncertainties, the model gives a good description of $A=3$ data.

%%%%%%%%%%%%%%%%%%%%%%%%%%%%%%%%%%%%%%%%%%%%%%%%%%%%%%% 
%%%%%%%%%%%%%%%%%%%%%%%%%%%%%%%%%%%%%%%%%%%%%%%%%%%%%%% 
%
%%%%%%%%%%%%%%%%%%%%%%%%%%%%%%%%%%%%%%%%%%%%%%%%%%%%%%% 
\begin{figure}[t]
\centering
\includegraphics[width=0.94\textwidth]{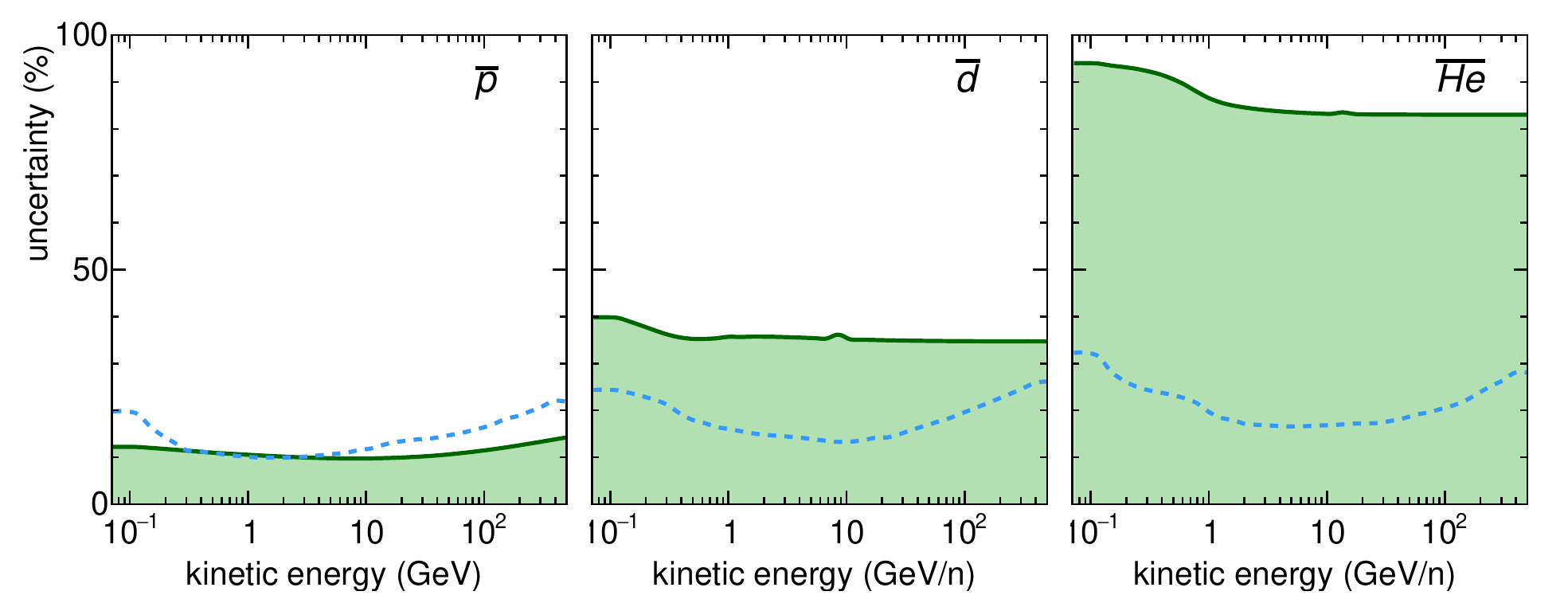}
\caption{\captionsize% 
  Relative uncertainties in the fluxes of \pbar, \dbar, and \hebar{} arising from production and destruction cross-sections.
  For comparisons, astrophysical uncertainties from CR injection and propagation are shown as dashed line.  
} 
\label{Fig::ccAbarAstroVSNuclear} 
\end{figure} 
%%%%%%%%%%%%%%%%%%%%%%%%%%%%%%%%%%%%%%%%%%%%%%%%%%%%%%% 
%
%
Along with \pbar{} production during DSA, we have also to account for their destruction,
as well as for the presence of non-annihilating reaction processes in the sources, \eg, \pbar+\p$\rightarrow$\pbar+X.
This process produces a ``tertiary'' DSA-accelerated antiproton component which can be important in the low-energy region. 
The total inelastic cross section $\sigma_{\rm in}(s)$ for \p-\p{} collisions is parameterised using
the formulae proposed by the PDG group, and for proton-nucleus and nucleus-nucleus collision is 
calculated using Tripathi universal parameterisation \citep{InelasticXS}.
We have tested different parameterisation such as those proposed in \citet{Moskalenko2002}.
To constrain the impact of destruction cross-section uncertainty in the flux calculations,
we adopt the method used in Carlson et al. \citep{Cirelli2014}, which consists in varying
the annihilating and non-annihilating branching ratios in order to evaluate the variation in the resulting fluxes.
In addition, we account for the production of tertiary particles.
The impact of these uncertainties in the resulting fluxes, however, depends on the underlying propagation model.
More details on this aspect will be discussed in a future paper.

%%%%%%%%%%%%%%%%%%%%%%%%%%%%%%%%%%%%%
\section{Results and discussions} %%%
\label{Sec::ResultsPbarBC}        %%%
%%%%%%%%%%%%%%%%%%%%%%%%%%%%%%%%%%%%%

Within the framework presented here, we have calculated the astrophysical background of \pbar, \dbar, and \hebar{} fluxes in CRs
together with their \emph{nuclear uncertainties}, \ie, those sources of systematic uncertainty linked to cross-section calculations.
Propagation calculations have been carried out analytically within the CR diffusion model in \citep{Feng2016}. 
The corresponding uncertainties are presented in Fig.\,\ref{Fig::ccAbarAstroVSNuclear} as solid green lines in terms of relative errors.
These lines represent the half-size of the uncertainty band in the resulting flux near-Earth, 
estimated from cross-section calculations of  Sect.\,\ref{Sec::Calculations},
then divided by the total flux.
For sake of comparisons, we also report preliminary estimates of \emph{astrophysical uncertainties} as blue dashed lines.
These uncertainties are related to the modeling of CR injection and transport processes in the Galaxy,
as well as to solar modulation of CRs in the Heliosphere.
Thanks to an increase in the accuracy of CR spectra and nuclear composition measurements of recent experiments,
and most notably in the \BC{} ratio, astrophysical uncertainties have seen a dramatic improvements in recent years.
On the other hand, the constraints provided to the coalescence mechanism
did not see a similar improvement, because they are still based on old data.
For \dbar{} and \hebar, it can be seen that nuclear uncertainties are dominating the total estimated  error in the calculated fluxes.
Our error estimation for \hebar{} is even optimistic, because it is directly extrapolated from \dbar{} cross-section data,
\ie, it is based on the assumption that the same coalescence momentum can be applied for \dbar{} and \tbar-\hebar{}.
Hence the production of \hebar{} and \tbar{} triplets, in these calculations, is directly extrapolated from
the constraints on the \dbar{} production cross-sections.
From Fig.\,\ref{Fig::ccHebarXS}, it can be seen that the existing data on \hebar{} and \tbar{} production support this assumption:
the \dbar-driven coalescence model output gives a fairly good description of the data. 
Further improvements may come from the ALICE experiment, which has collected
precious antinuclei production data from \p-\p{} collisions at LHC.
\\
\\
{\footnotesize%
We thank our colleagues of the AMS Collaboration for valuable discussions.
%NT acknowledges the European Commission for support under the H2020-MSCA-IF-2015 action, grant agreement No.707543-MAtISSE.
AO acknowledges CIEMAT, CDTI and SEIDI MINECO under grants ESP2015-71662-C2-(1-P) and MDM-2015-0509.
NT acknowledges support from MAtISSE.
This project has received funding from the European Union's Horizon 2020 research and innovation programme under the Marie Sklodowska-Curie grant agreement No 707543.
}

%%%%%%%%%% BIBLIOGRAPHY %%%%%%%%%%%%%%%%%%%%%%%%%%%%%%%


\begin{thebibliography}{99}
%%%%%%%%%%%%%%%%%%%%%%%%%%%%%%%%%%%%%%%%%%%%%%%%%%%%%%%

\bibitem{Aramaki2016} 
T. Aramaki, \etal, %S. Boggs, S. Bufalino, \etal,
\href{http://dx.doi.org/10.1016/j.physrep.2016.01.002}
{Phys. Rept. 618, 1-38 (2016)};
%
L. A. Dal, \& M. Kachelrie\ss,
\href{http://dx.doi.org/10.1103/PhysRevD.86.103536}
{Phys. Rev. D 86, 103536 (2012)};
%
A. Ibarra, \& S. Wild, %arXiv:1301.3820
\href{http://dx.doi.org/10.1103/PhysRevD.88.023014}
{Phys. Rev. D 88, 023014 (2013)};
%
L. A. Dal, \& A. R. Raklev, %1504.07242
\href{http://dx.doi.org/10.1103/PhysRevD.91.123536}
{Phys. Rev. D 91, 123536 (2015)}.

\bibitem{Cirelli2014}
%
E. Carlson, \etal, %A. Coogan, T. Linden, S. Profumo, A. Ibarra, S. Wild,
\href{http://dx.doi.org/10.1103/PhysRevD.89.076005}
{Phys. Rev. D 89, 076005 (2014)};
%
M. Cirelli, \etal, %N. Fornengo, M. Taosoa,  A. Vittino,
\href{http://dx.doi.org/10.1007/JHEP08(2014)009}
{JHEP 08, 009 (2014)};
%
K. Blum, \etal, %K. C. Y. Ng, R. Sato, M. Takimoto,
[\href{https://arxiv.org/abs/1704.05431}
{arXiv:1704.05431}] (2017);
%
A. Coogan \& S. Profumo, [\href{https://arxiv.org/abs/1705.09664}
{arXiv:1705.09664}] (2017).
%
%J. Sokol, {Science, April 4$^{\rm th}$, 2017} [\href{http://dx.doi.org/10.1126/science.aal1067}{ScienceMag}]

\bibitem{Aguilar2016PbarP}
M. Aguilar, \etal,
\href{http://dx.doi.org/10.1103/PhysRevLett.117.091103}
{Phys. Rev. Lett. 117, 091103 (2016)}.

\bibitem{Aguilar2016BC}
M. Aguilar, \etal, 
\href{https://dx.doi.org/10.1103/PhysRevLett.117.231102}
{Phys. Rev. Lett. 117, 231102 (2016)}.

\bibitem{TomassettiDonato2015}
N. Tomassetti \& F. Donato,
\href{http://dx.doi.org/10.1088/2041-8205/803/2/L15}
{Astrophys. J. 803, L15 (2015)};
%
M. Kachelrie\ss, A. Neronov, \& D. V. Semikoz,
\href{http://dx.doi.org/10.1103/PhysRevLett.115.181103}
{Phys. Rev. Lett. 115, 181103 (2015)};
%
N. Tomassetti,
\href{http://dx.doi.org/10.1088/2041-8205/815/1/L1}
{Astrophys. J. 815, L1 (2015)};
%
Y. Ohira, N. Kawanaka, \& K. Yoka,
\href{https://doi.org/10.1103/PhysRevD.93.083001}
{Phys. Rev. D 93, 083001 (2016)}.

\bibitem{PbarSnr}
P. Blasi, \& P. D. Serpico,
\href{http://dx.doi.org/10.1103/PhysRevLett.103.081103}
{Phys. Rev. Lett. 103, 081103 (2009)};
%
P. Mertsch, \& S. Sarkar,
\href{http://dx.doi.org/10.1103/PhysRevD.90.061301}
{Phys. Rev. D 90, 061301 (2014)};
%
N. Tomassetti, \& J. Feng,
\href{http://dx.doi.org/10.3847/2041-8213/835/2/L26}
{Astrophys. J. 835, L26 (2017)}.

\bibitem{Giesen2015}
G. Giesen, \etal,
\href{http://dx.doi.org/10.1088/1475-7516/2015/09/023}
{JCAP 09, 23 (2015)};
%
C. Evoli, D. Gaggero, \& D. Grasso,
\href{http://dx.doi.org/10.1088/1475-7516/2015/12/039}
{JCAP 12, 039 (2015)}.

\bibitem{Feng2016}
J. Feng, N. Tomassetti, \& A. Oliva,
\href{http://dx.doi.org/10.1103/PhysRevD.94.123007}
{Phys. Rev. D 94, 123007 (2016)};
%
N. Tomassetti,
\href{http://dx.doi.org/10.1088/2041-8205/752/1/L13}
{Astrophys. J. 752, L13 (2012)};
%
N. Tomassetti,
\href{http://dx.doi.org/10.1103/PhysRevD.92.081301}
{Phys. Rev. D 92, 081301 (2015)}. %(R)

\bibitem{Grenier2015} 
I. A. Grenier, J. H. Black, \& A. W. Strong,
\href{http://dx.doi.org/10.1146/annurev-astro-082214-122457}
{Annu. Rev. Astron. Astrophys., 53, 199–246 (2015)};
%
A. W. Strong, I. V. Moskalenko, \& V. S. Ptuskin,
\href{http://dx.doi.org/10.1146/annurev.nucl.57.090506.123011}
{Ann. Rev. Nucl. \& Part. Sci. 57 285--327 (2007)};
%
E. Amato \& P. Blasi, 
\href{https://doi.org/10.1016/j.asr.2017.04.019}
{Adv. Space Res., in press (2017)}
[\href{https://arxiv.org/abs/1704.05696}{arXiv:1704.05696}];
%
P. Blasi, 
\href{http://dx.doi.org/10.1007/s00159-013-0070-7}
{Astron. Astrophys. Rev. 21, 70 (2013)}.

\bibitem{TanNg:1983}
L. C. Tan, \& L. K. Ng,
\href{http://dx.doi.org/10.1088/0305-4616/9/10/015/}
{J. Phys. G: Nucl. Phys. 9, 1289 (1983)};
%
A. N. Kalinovskyi, \etal, % N. V. Mokhov, Yu. P. Nikitin, 
AIP, New-York (1989).

\bibitem{diMauro:2014zea}
M. di Mauro, \etal, %F. Donato, A. Goudelis, P. D. Serpico, 
\href{http://dx.doi.org/10.1103/PhysRevD.90.085017}
{Phys. Rev. D 90, 085017 (2014)}.

\bibitem{Kachelriess:2015wpa}
M. Kachelrie\ss, I. Moskalenko, \& S. S. Ostapchenko,
\href{http://dx.doi.org/10.1088/0004-637X/803/2/54}
{Astrophys. J. 803, 54 (2015)}.

\bibitem{Kappl:2014}
R. Kappl \& M. W. Winkler,
\href{http://dx.doi.org/10.1088/1475-7516/2014/09/051}
{JCAP 09, 51 (2014)}.

\bibitem{NA49}
T. Anticic, \etal,
\href{http://dx.doi.org/10.1140/epjc/s10052-009-1172-2}
{Eur. Phys. J. C 65, 9-63 (2010)};
%
B. Baatar, \etal,
\href{http://dx.doi.org/10.1140/epjc/s10052-013-2364-3}
{Eur. Phys. J. C 73, 2364 (2013)}.

\bibitem{Arsene:2007jd}
I. Arsene, \etal,
\href{http://dx.doi.org/10.1103/PhysRevLett.98.252001}
{Phys. Rev. Lett. 98, 252001 (2007)}.

\bibitem{Aamodt:2011zj} 
K. Aamodt, \etal,
\href{http://dx.doi.org/10.1140/epjc/s10052-011-1655-9}
{Eur. Phys. J. C 71, 1655 (2011)}.

\bibitem{Pierog:2013ria}
T. Pierog, \etal,
\href{http://dx.doi.org/10.1103/PhysRevC.92.034906}
{Phys. Rev. C. 92, 034906 (2015)}.

\bibitem{Engel:1999db}
R. Engel, \etal, %T. K. Gaisser, T. Stanev, P. Lipari,
Proc. 26$^{\rm th}$ ICRC (Salt Lake City, USA),
\href{https://inspirehep.net/record/512007}
{Vol 1, 415-418 (1999)}.

\bibitem{Ostapchenko:2010vb}
S. Ostapchenko,
\href{http://dx.doi.org/10.1103/PhysRevD.83.014018}
{Phys. Rev. D 83, 014018 (2011)}.

\bibitem{Fisher2003} 
H. G. Fisher,
\href{http://dx.doi.org/10.1556/APH.17.2003.2-4.20}
{APH N.S., Heavy Ion Phys. 17, 369-386}.

\bibitem{Donato2008}
P. Chardonnet, J. Orloff, \& P. Salati,
\href{http://dx.doi.org/10.1016/S0370-2693(97)00870-8}
{Phys. Lett. B 409, 3, 13-320 (1997)};
%
F. Donato, N. Fornengo, \& D. Maurin,
\href{http://dx.doi.org/10.1103/PhysRevD.78.043506}
{Phys. Rev. D 78, 043506 (2008)};
%
R. Duperray, \etal, %B. Baret, D. Maurin, ...
\href{http://dx.doi.org/10.1103/PhysRevD.71.083013}
{Phys. Rev. D 71, 083013 (2005)}.

\bibitem{InelasticXS}	
J. Beringer \etal,  
\href{http://dx.doi.org/10.1103/PhysRevD.86.010001}
{Phys. Rev. D 86, 010001 (2012)};
%
R. K. Tripathi, M. G. Cucinotta, \& J. W. Wilson,
\href{http://dx.doi.org/10.1016/0168-583X(96)00331-X}
{Nucl. Inst. Meth. B 117, 347-349 (1996)}.

\bibitem{Moskalenko2002} 
I. V. Moskalenko, \etal, %A. W. Strong, J. F. Ormes, M. S. Potgieter,
\href{http://dx.doi.org/10.1086/324402}
{\ApJ{} 565, 280 (2002)};
%
A. A. Moiseev, \& J. F. Ormes,
\href{https://doi.org/10.1016/S0927-6505(96)00071-0}
{Astropart. Phys. 6, 379-386 (1997)};
%
V. F. Kuzichev, Yu. B. Lepikhin, \& V. A. Smirnitsky,
\href{https://doi.org/10.1016/0375-9474(94)90745-5}
{Nucl. Phys. A 576, 581 (1994)}.


\bibitem{DbarXS} % British-Scandinavian-MIT at CERN/ISR
B. Alper, \etal, %H. Boeggild, P. S. L. Booth, \etal,
\href{http://dx.doi.org/10.1016/0370-2693(73)90730-2}
{Phys.Lett. 47B (1973) 275-280};
%
W. M. Gibson, \etal, %A. Duane, H. Newman, \etal, 
\href{http://dx.doi.org/10.1007/BF02822248}
{Lett. Nuovo Cim. 21, 189 (1978)}.

\end{thebibliography}
\end{document}